\documentclass[10pt,twocolumn,letterpaper]{article}

\usepackage{cvpr}              

\usepackage[accsupp]{axessibility}
\usepackage{graphicx}
\usepackage{booktabs}

\usepackage{cite}
\usepackage{amsmath,amssymb,amsfonts}

\DeclareMathOperator*{\argmin}{arg\,min}
\usepackage{algorithmic}

\usepackage{textcomp}
\usepackage{epsfig}
\usepackage{amsmath,amssymb,mathtools}
\usepackage{rotating,multirow,array,booktabs}
\usepackage{comment}
\usepackage{hhline}
\usepackage{xcolor}
\usepackage{url}
\usepackage[ruled,vlined]{algorithm2e}
\usepackage{lipsum}
\usepackage{bm}
\usepackage{setspace}
\usepackage{makecell}
\usepackage{cellspace}
\usepackage{pifont}
\usepackage{float}
\usepackage{siunitx}
\usepackage{array}
\newcolumntype{P}[1]{>{\centering\arraybackslash}p{#1}}
\newcolumntype{M}[1]{>{\centering\arraybackslash}m{#1}}

\usepackage[section]{placeins}

\definecolor{red}{rgb}{0.7,0,0}

\definecolor{pink}{rgb}{1,0.03,0.5}

\usepackage{lipsum}
\newcommand\blfootnote[1]{
\begingroup
\renewcommand\thefootnote{}\footnote{#1}
\addtocounter{footnote}{-1}
\endgroup
}

\def\@fnsymbol#1{\ensuremath{\ifcase#1\or *\or \dagger\or \ddagger\or
   \mathsection\or \mathparagraph\or \|\or **\or \dagger\dagger
   \or \ddagger\ddagger \else\@ctrerr\fi}}
\newcommand{\ssymbol}[1]{^{\@fnsymbol{#1}}}

\newcommand*{\affaddr}[1]{#1} 
\newcommand*{\affmark}[1][*]{\textsuperscript{#1}}

\usepackage[pagebackref,breaklinks,colorlinks]{hyperref}

\usepackage[capitalize]{cleveref}
\crefname{section}{Sec.}{Secs.}
\Crefname{section}{Section}{Sections}
\Crefname{table}{Table}{Tables}
\crefname{table}{Tab.}{Tabs.}

\begin{document}

\title{Learning to Deblur using Light Field Generated and Real Defocus Images}
\author{
Lingyan Ruan\affmark[1]$\ssymbol{1}$\space\space\space\space\space\space\space\space   Bin Chen\affmark[2]$\ssymbol{1}$\space\space\space\space\space\space\space\space      Jizhou Li\affmark[3]\space\space\space\space\space\space\space\space     Miuling Lam\affmark[1]$\ssymbol{2}$\\
\affaddr{\affmark[1]City University of Hong Kong}\space\space\space\space\space\space 
\affaddr{\affmark[2]Max-Planck-Institut für Informatik}\space\space\space\space\space\space 
\affaddr{\affmark[3]Stanford University}\\
\affaddr{\href{http://lyruan.com/Projects/DRBNet}{\normalsize{http://lyruan.com/Projects/DRBNet}}}
}
\maketitle
\blfootnote{$*$ denotes equal contribution and $\dagger$ denotes corresponding author.}

\begin{abstract}
Defocus deblurring is a challenging task due to the spatially varying nature of defocus blur. While deep learning approach shows great promise in solving image restoration problems, defocus deblurring demands accurate training data that consists of all-in-focus and defocus image pairs, which is difficult to collect. Naive two-shot capturing cannot achieve pixel-wise correspondence between the defocused and all-in-focus image pairs. Synthetic aperture of light fields is suggested to be a more reliable way to generate accurate image pairs. However, the defocus blur generated from light field data is different from that of the images captured with a traditional digital camera. In this paper, we propose a novel deep defocus deblurring network that leverages the strength and overcomes the shortcoming of light fields. We first train the network on a light field-generated dataset for its highly accurate image correspondence. Then, we fine-tune the network using feature loss on another dataset collected by the two-shot method to alleviate the differences between the defocus blur exists in the two domains. This strategy is proved to be highly effective and able to achieve the state-of-the-art performance both quantitatively and qualitatively on multiple test sets. Extensive ablation studies have been conducted to analyze the effect of each network module to the final performance.

\end{abstract}

\section{Introduction}
\label{sec:intro}
\noindent{The} use of large camera aperture can increase the luminous flux so that the image can be captured with a shorter exposure time. However, this also reduces the depth of field (DOF) - only points near the focal plane can be captured sharply, while a point far from the focal plane will cast to the camera sensor, instead of a single image point, a spot called the Circle of Confusion (COC) \cite{potmesil1981lens} and result in defocus blur. Shallow DOF is sometimes an aesthetic effect pursued sedulously by the photographer \cite{hach2015cinematic,sakurikar2018refocusgan}, but it may also degrade important visual information. Thus, restoring an all-in-focus image from its defocused version is highly demanded to reveal the latent information and benefit to artificial intelligence applications such as object detection \cite{redmon2018yolov3} and text recognition \cite{lyu2018mask}. Despite its great potential, defocus deblurring remains a challenging problem due to its spatially varying nature - every point has its own diameter of COC depending on the depth of the corresponding scene point. Besides, the shape of COC varies with respect to the relative position from the optical axis. To address defocus blur, the most intuitive way is to first estimate the blur kernel for each pixel, then apply non-blind deconvolution \cite{levin2007image,karaali2017edge,shi2015just,d2016non,park2017unified,lee2019deep}. However, both steps have limitations. First, blur kernel estimation is not accurate and is often based on simple Gaussian \cite{lee2019deep,park2017unified,karaali2017edge} or disk kernel \cite{d2016non} assumption. Second, deconvolution tends to introduce ringing artifact on edges due to the Gibbs phenomenon \cite{yuan2007image} even if an accurate blur kernel is given. \\
\begin{figure}[t]
    \setlength{\abovecaptionskip}{0.1cm}
    \setlength{\belowcaptionskip}{0.1cm}
	\centering
	\includegraphics[width=0.9\linewidth]{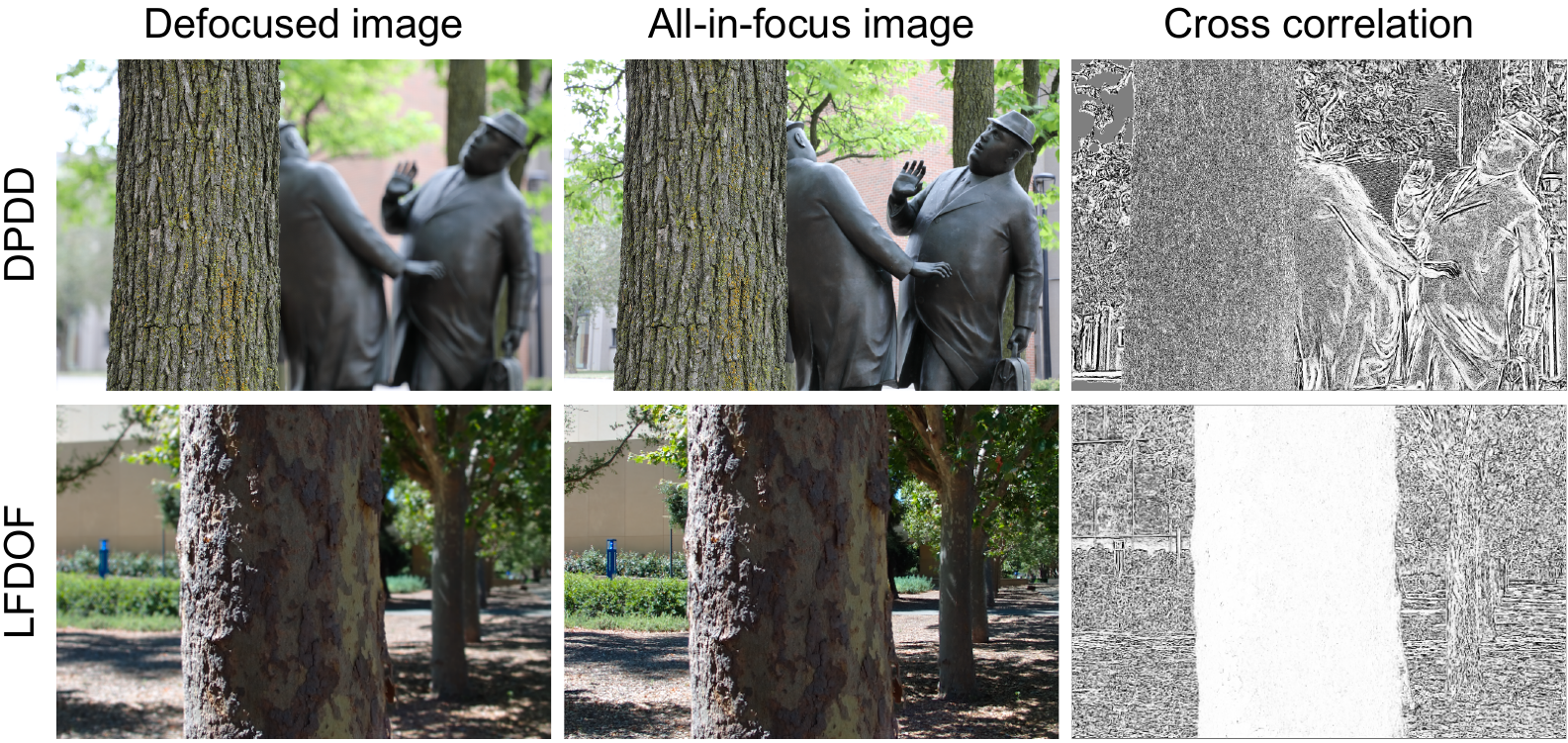}
	\caption{Cross correlation between defocused and all-in-focus image pair in DPDD \cite{abuolaim2020defocus} (top) and LFDOF dataset \cite{ruan2021aifnet} (bottom).}
	\label{fig:cross_correlation}
	\vspace{-0.05cm}
\end{figure}
Recently, researchers have adopted end-to-end deep neural networks to directly restore sharp images from defocus blur \cite{abuolaim2020defocus,lee2021iterative,son2021single}, which largely outperform the conventional two-step approaches in terms of performance and efficiency. These networks are all trained on a dataset called Dual-Pixel Defocus Deblurring (DPDD) \cite{abuolaim2020defocus} which is captured sequentially with different aperture sizes to attain defocused and all-in-focus image pairs. However, it is hardly possible to capture defocused and all-in-focus pairs with accurate correspondence in two shots, especially for outdoor scenes due to moving objects (\eg, plants, cars) and illuminance variation.
To this end, another dataset LFDOF \cite{ruan2021aifnet} is built utilizing the benefit of light field refocusing and synthetic aperture to generate a large number of defocused images with a variety of DOFs and focal distances from a single light field sample. To examine the consistency among the image pairs, we select similar scenes from the two datasets and calculate the cross correlation between defocused and all-in-focus pairs. As shown in Fig~\ref{fig:cross_correlation}, LFDOF has strong cross correlation in the sharp regions, whereas DPDD does not hold consistence even at the sharp regions (the tree trunk is in focus in both defocused and all-in-focus images). However, despite good pixel-wise consistency of LFDOF, the defocus blur generated from light field data is not the same as that captured with a conventional digital camera (see Sec. \ref{sec:dslr_vs_lf}).  
In this paper, we intend to make full use of the advantages of LFDOF and DPDD datasets to train a deep network for defocus deblurring.
In summary, the contributions of this paper are as follows:
\begin{itemize}
\setlength\itemsep{0.01cm}
  \item We analyze the characteristics of two defocus blur datasets LFDOF and DPDD and develop a novel training strategy for single image defocus deblurring. We also estimate and compare the Point Spread Function (PSF) of light field generated defocus blur against that captured with a conventional digital camera.
  \item We propose an end-to-end network architecture equipped with a novel dynamic residual block to reconstruct the sharp image in a coarse-to-fine manner.
  \item We conduct extensive experiments to evaluate the effect of each network module and demonstrate the state-of-the-art performance quantitatively and qualitatively on multiple test sets.   
\end{itemize}
\section{Related Work}
\label{sec:related_work}
\paragraph{Conventional Methods} Conventional defocus deblurring methods typically follow a two-step approach consisting of defocus map (blur level for each pixel) estimation \cite{tai2009single,zhuo2011defocus,karaali2017edge} followed by non-blind deconvolution \cite{krishnan2009fast,levin2007image,fish1995blind}. Much effort has been made to improve the accuracy of defocus map as it influences the deblurring performance significantly \cite{shi2015just,karaali2017edge,shi2015just,park2017unified}. 
However, this approach usually requires intensive computation while ending up with limited performance due to the defective intermediate defocus map.  

\vspace{0.08cm}
\noindent \textbf{Defocus Blur Dataset} { } There are several publicly available datasets for problems related to defocus blur. Defocus blur datasets comprising real defocused RGB images and binary masks are built by Shi \etal \cite{shi2015just} and Zhao \etal \cite{zhao2019defocus}. But they can only work for blur detection due to the absence of all-in-focus images. Abuolaim and Brown \cite{abuolaim2020defocus} built the DPDD dataset using a dual-pixel camera and capturing the defocus and all-in-focus pairs in two successive shots. Ruan \etal \cite{ruan2021aifnet} proposed the light field based defocus deblurring dataset LFDOF leveraging the synthetic aperture and refocusing features of light field technology \cite{ng2005light}, where the image pairs were acquired in single shots.
Lee \etal \cite{lee2021iterative} provided a benchmark test set consisting of 50 scenes captured with a dual-camera system with beam splitter. Adopting existing RGBD dataset \cite{silberman2012indoor,ros2016synthia} to generate defocus images based on the depth maps, Lee \etal \cite{lee2019deep} have simulated single defocused images while Abuolaim \etal \cite{abuolaim2021learning} and Pan \etal \cite{pan2021dual} have simulated dual defocused pairs. However, the images in these datasets are synthetic, thus lacking realism. In this paper, we use the DPDD and LFDOF datasets that are captured in real-world scenes.

\vspace{0.08cm}
\noindent \textbf{CNN-based Methods} { } Abuolaim and Brown \cite{abuolaim2020defocus} adopted an U-Net-like architecture to restore the sharp image in an end-to-end manner and their follow-up work  \cite{abuolaim2022improving} incorporated a single-encoder multi-decoder architecture to further improve the performance. Lee \etal \cite{lee2021iterative} proposed a network equipped with iterative filter adaptive module to tackle the spatially varying defocus blur and auxiliary reblurring module to enhance the restoration performance. Son \etal \cite{son2021single} proposed an effective way to simulate inverse kernels via kernel sharing parallel atrous convolution block.  The aforementioned networks are all trained on DPDD dataset \cite{abuolaim2020defocus}. Ruan \etal \cite{ruan2021aifnet} addressed single image defocus deblurring based on the conventional two-step strategy that takes the intermediate defocus map as the guidance for the deblurring step. Some other works used dual views to address the defocus deblurring problem \cite{vo2021attention,pan2021dual}. However, we focus on single image defocus deblurring in this paper.

\vspace{0.08cm}
\noindent \textbf{Dynamic Filtering} { } Dynamic filtering, also called filter adaptive convolution, has been successfully adopted in various low level vision tasks, such as denoising \cite{mildenhall2018burst}, video deblurring \cite{zhou2019spatio}, super-resolution \cite{jo2018deep,kim2021koalanet,xu2020unified} and defocus deblurring \cite{lee2021iterative}, \etc, since first introduced by Jia \etal  \cite{jia2016dynamic}. It aims to learn per-pixel kernels instead of a single kernel over the entire image and thus is capable of handling non-uniform or spatially varying degradation. 
Our proposed dynamic residual block is partially inspired by dynamic filtering, but instead of directly applying convolution with the learned per-pixel kernels to get the output, we treat them as the learned dynamic residuals along with the direct residual for better performance. Also, being different from the method of Lee \etal \cite{lee2021iterative} which transforms the features with the learned kernels to remove the spatially varying blur in the feature space, we learn the dynamic residual in the image domain in a coarse-to-fine manner to restore the sharp image. Section \ref{subsec:Analysis_and_Discussion} demonstrates the superior performance of our proposed method against existing works in addressing single image defocus deblurring.
\section{Conventional Digital Camera \vs Light Field Camera}
\label{sec:dslr_vs_lf}
\begin{figure*}[ht]
    \setlength{\abovecaptionskip}{0.1cm}
    \setlength{\belowcaptionskip}{0.1cm}
	\centering
	\includegraphics[width=\linewidth]{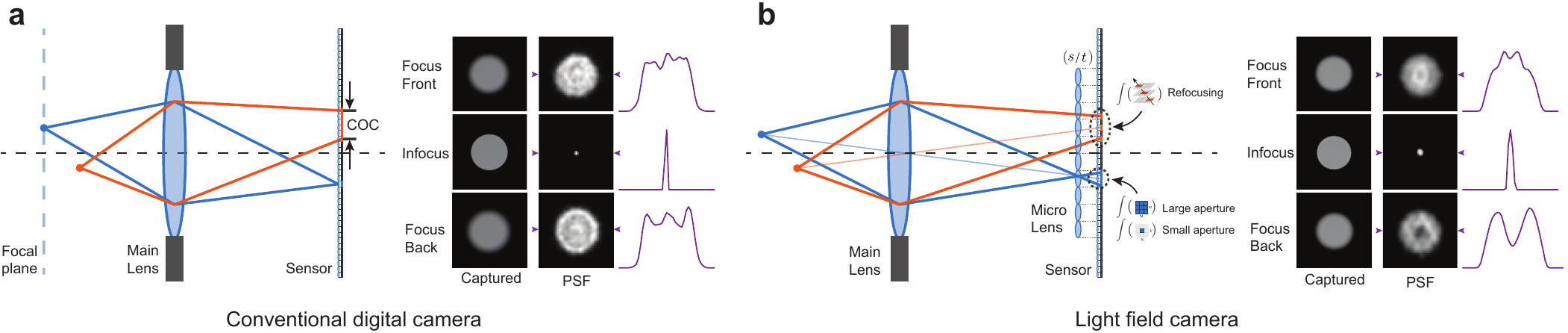}
	\caption{The defocus blur formation process of: (a) conventional digital camera (Canon EOS R5), and (b) Light field camera (Lytro Illum). The captured disk pattern, estimated PSF, and center line profile of PSF image are shown alongside the imaging optical path diagram. The diagrams of light field synthetic aperture and refocusing process are shown in (b).}  
	\label{fig:dslr_lf}
	\vspace{-0.5cm}
\end{figure*}
\noindent We begin with describing the defocus blur discrepancy between the images captured with conventional digital cameras (Canon EOS R5) and light field cameras (Lytro Illum) respectively as shown in Fig.~\ref{fig:dslr_lf}.

For a conventional digital camera (Fig.~\ref{fig:dslr_lf}a), the rays emitted from a scene point on the focal plane converge to a single pixel of the image sensor by the main lens, while a point away from the focal plane projects to a patch of pixels on the sensor in a circular shape (COC), causing defocus blur. For a light field camera, a micro lens array is placed in front of the sensor, thus the rays coming from the main lens are re-distributed to the pixels under micro lenses, which means that each pixel does not only record the integrated illuminance but also the directional information of the rays. 
Each sub-aperture view of a light field only records a small part ($\frac{1}{14\times14}$ for Lytro Illum) of the full aperture. After capturing, the aperture size can be further synthesized by integrating an appropriate subset of samples from multiple sub-aperture views. Similarly, refocusing to different depth can be achieved by integrating pixels along different directions on the epipolar plane image (EPI) \cite{bolles1987epipolar} as shown in Fig. \ref{fig:dslr_lf}b. 

To understand the differences between the defocus blur produced by these two types of cameras, we estimate and visualize their PSFs using the algorithm proposed by Mannan and Langer \cite{mannan2016blur}. The PSFs of three typical cases: front focus, in focus, and back focus are estimated and shown alongside the imaging optical path diagram in Fig.~\ref{fig:dslr_lf}. Generally, the PSFs produced by the digital camera follows the diffraction pattern of single Airy disk, while the PSFs produced by the light field camera resemble the patterns of multiple Airy disks, which can be explained by the synthetic nature of light field generated defocus blur. More PSF estimations can be found in the supplementary materials.

\section{Methodology}
\label{sec:method}

\begin{figure*}[htb]
    \setlength{\abovecaptionskip}{0.1cm}
    \setlength{\belowcaptionskip}{0.1cm}
	\centering
	\includegraphics[width=\linewidth]{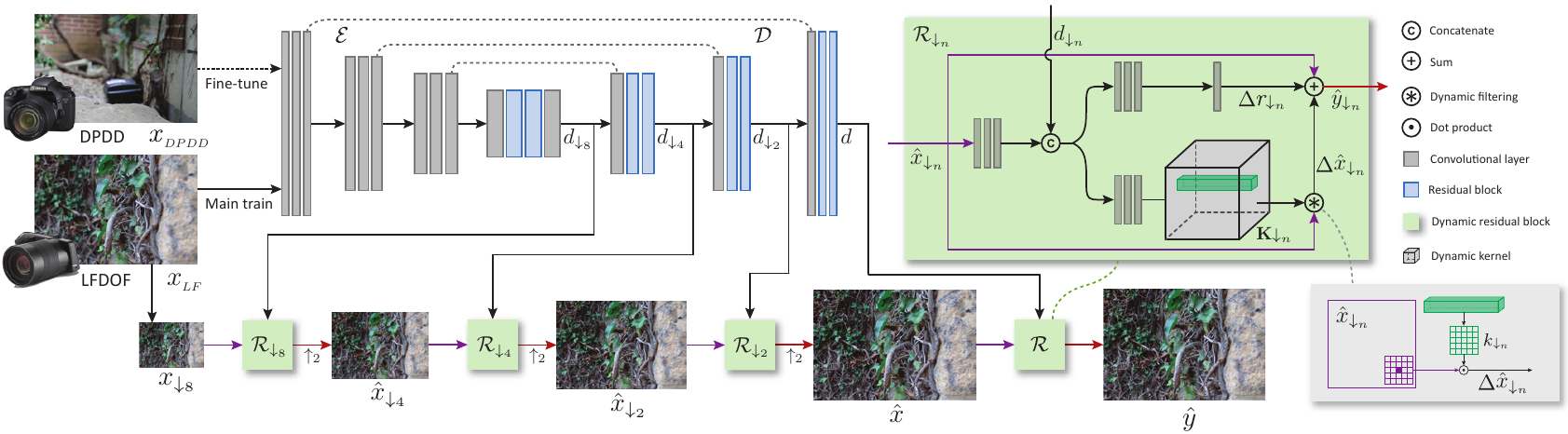}
	\caption{Our network architecture is mainly composed of encoder $\mathcal{E}$, decoder $\mathcal{D}$ and four dynamic residual modules $\mathcal{R}$. The LFDOF dataset is used for main training and the DPDD dataset is used for network fine-tuning.}   
	\label{fig:network_structure}
	\vspace{-0.4cm}
\end{figure*}

\noindent Single image defocus deblurring aims to recover the latent sharp image $\hat{y}$ from an observed input $x$ that is distorted by defocus blur. A deep network can be trained as a mapping function $\mathcal{F}$ parameterized by $\theta$:
\vspace{-0.1cm}
\begin{equation}
    \hat{y} = \mathcal{F}_\theta(x)
\end{equation}
A loss function should be tailored to optimize $\theta$ in order to minimize the distance between $\hat{y}$ and $y$:
\begin{equation}
    \hat{\theta} = \argmin_{\theta}\sum\nolimits_{i}{\mathcal{L}(\mathcal{F}_\theta (x_i), y_i)},
\end{equation}
where ($x_i$, $y_i$) are defocused and all-in-focus image pairs.
As shown in Fig.~\ref{fig:network_structure}, we design our network $\mathcal{F}$ in an encoder-decoder \cite{ronneberger2015u} structure. The encoder ($\mathcal{E}$) extracts the multi-scale pyramidal features, which are then added to the corresponding scale of decoder ($\mathcal{D}$) by skip connections to stabilize the network training. Two residual blocks \cite{he2016deep} are added to each scale of decoder.
We will show in Sec. \ref{subsec:Analysis_and_Discussion} that simple encoder-decoder structure cannot handle defocus deblurring well. 

\begin{figure}[htb]
    \setlength{\abovecaptionskip}{0.1cm}
    \setlength{\belowcaptionskip}{0.1cm}
	\centering
	\includegraphics[width=.95\linewidth]{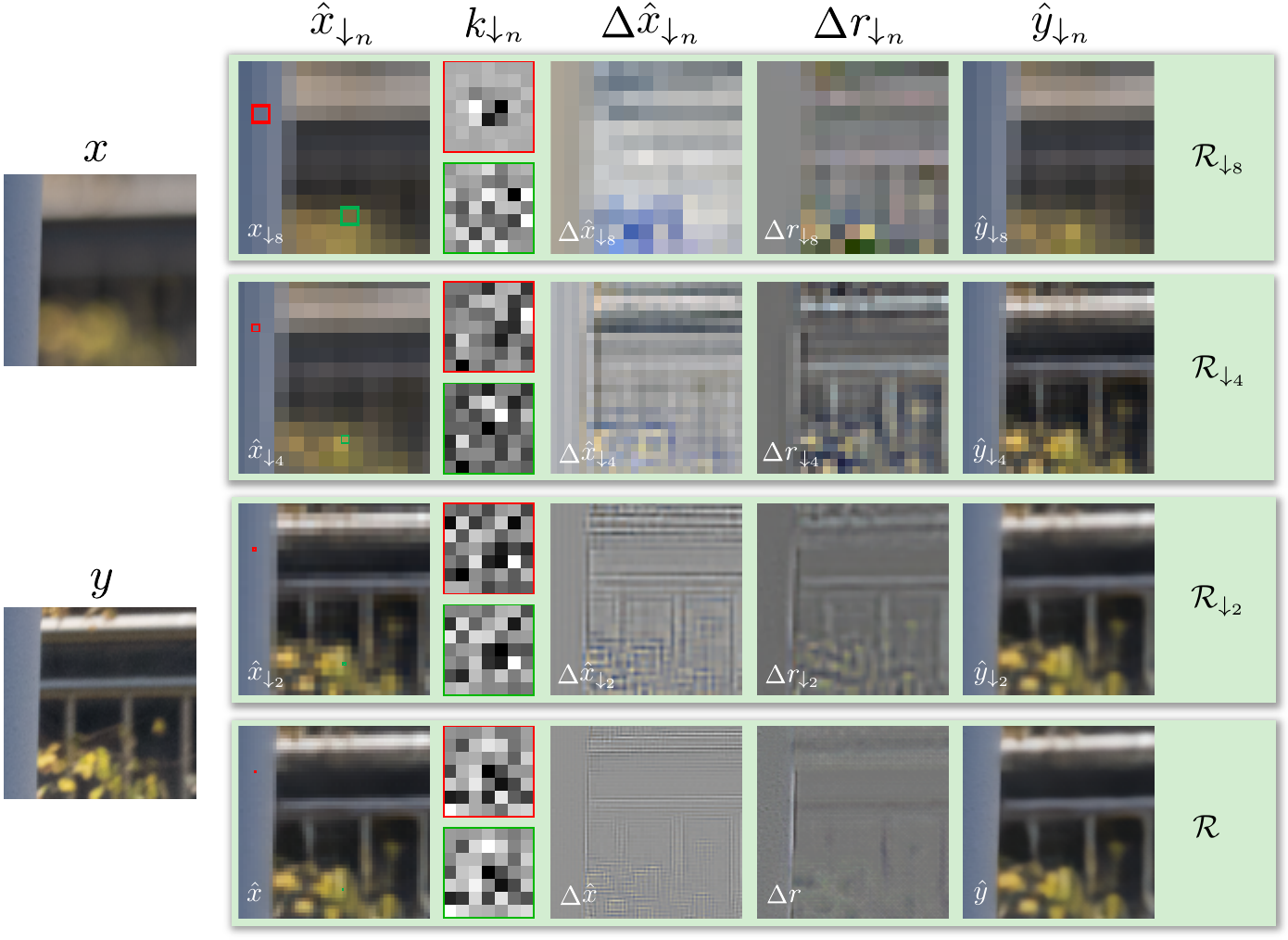}
	\caption{Visualization of each step in DRB on all scales. From left to right: original input $x$ and ground truth $y$, input of DRB $\hat{x}_{\downarrow_n}$, dynamic kernel on two pixels $k_{\downarrow_n}$, dynamic residual $\Delta \hat{x}_{\downarrow_n}$, residual $\Delta r{\downarrow_n}$, output of DRB $\hat{y}_{\downarrow_n}$.}
	\label{fig:drb}
    \vspace{-0.2cm}
\end{figure}

\vspace{0.08cm}
\noindent \textbf{Dynamic Residual Block (DRB)} { } Inspired by the dynamic filtering approach \cite{jia2016dynamic, lee2021iterative,jo2018deep,xu2020unified,sim2019deep}, we design a residual version of dynamic filtering block $\mathcal{R}$ to better handle the spatially varying defocus blur. We connect the dynamic residual block to each scale of decoder in a cascaded fashion to restore the latent sharp image progressively. Each dynamic residual block can be formulated as:
\begin{equation}
\hat{y}_{\downarrow_n} = \mathcal{R}(\hat{x}_{\downarrow_n}, d_{\downarrow_n}; \theta_r),
\end{equation}
where $\hat{x}_{\downarrow_n}$ represents one of the input of $\mathcal{R}$, which is also the $\times 2$ up-sampled version of the output from the previous dynamic residual block: $\hat{x}_{\downarrow_n}\,=\, \uparrow_2({\hat{y}_{\downarrow\frac{n}{2}}})$. Note that the input of the first dynamic residual block is $x_{\downarrow_8}$, which is a direct down-sample from the input image $x$.
Specifically, as illustrated in the green inset in Fig.~\ref{fig:network_structure}, the input $\hat{x}_{\downarrow_n}$ is passed to three convolution layers and concatenated with the equivalent size feature map $d_{\downarrow_n}$ from decoder $\mathcal{D}$, then sent to two paths: one is to estimate the dynamic kernel volume $\mathbf{K}$ and another is to estimate the residual $\Delta r_{\downarrow_n}$. The estimated dynamic kernel volume $\mathbf{K}$ is then convolved with the input $\hat{x}_{\downarrow_n}$ to obtain the dynamic residual $\Delta \hat{x}_{\downarrow_n}$:
\begin{equation}
\Delta \hat{x}_{\downarrow_n} = \hat{x}_{\downarrow_n} \circledast \mathbf{K}
\label{eq:dynamic_residual}
\end{equation}
The dynamic filtering procedure is depicted within the gray inset in Fig.~\ref{fig:network_structure}. Finally the output of each dynamic residual block can be calculated by:
\begin{equation}
\hat{y}_{\downarrow_n} = \hat{x}_{\downarrow_n} + \Delta r_{\downarrow_n} + \Delta \hat{x}_{\downarrow_n}, 
\label{eq:drb_output}
\end{equation}
We visualize a small patch in each step of our four dynamic residual blocks in Fig.~\ref{fig:drb}, which clearly shows how the latent sharp image is reconstructed step by step from its defocused version. Specifically, the dynamic residual $\Delta \hat{x}_{\downarrow_n}$ extracts the high frequency features like edges and corner points, which are lost during the defocus blur formation process, while the residual $\Delta r_{\downarrow_n}$ focus on the low frequency features that represent the essential content covered by defocus blur. The dynamic residual $\Delta \hat{x}_{\downarrow_n}$, residual $\Delta r_{\downarrow_n}$ and the input image $\hat{x}_{\downarrow_n}$ jointly contribute to recovering the latent sharp image. Section \ref{subsec:Analysis_and_Discussion} demonstrates the effectiveness of the proposed DRB. 

\noindent \textbf{Training Strategy and Loss} {} As demonstrated in Sec. \ref{sec:dslr_vs_lf}, the defocus blur produced by a conventional digital camera and a light field camera are different. To remedy this gap, we propose a training strategy that leverages the strength and overcomes the shortcoming of light field data. Specifically, we apply the light field generated dataset LFDOF in the main training round for its highly accurate image correspondence. Then, we fine-tune the network using DPDD dataset to alleviate the differences between the two domains.
Different losses are used in each stage. In the main training stage, $\ell_1$ norm is used:
\begin{equation}
\mathcal{L}_1^{\scalebox{.95}{$\scriptscriptstyle \textrm{LF}$}} = \|  \hat{y}_{\scalebox{.95}{$\scriptscriptstyle \textrm{LF}$}} - y_{\scalebox{.95}{$\scriptscriptstyle \textrm{LF}$}}\|_1\\
\label{eq:loss:lf}
\end{equation}
During the fine-tuning stage, pixel-wise loss should be avoided because of the misalignment (induced from two-shot) between the defocused and all-in-focus pairs in DPDD dataset. We apply the VGG-based feature loss \cite{johnson2016perceptual} in this step to transfer the learned knowledge to the target domain in the feature space, thus avoiding exact matching in the image space. We extract the feature maps from the 2nd, 7th, and 14th layer of the pre-trained VGG-19 network \cite{simonyan2014very}, denoted as $\phi$.
\begin{equation}
\mathcal{L}_{\scalebox{.95}{$\scriptscriptstyle \textrm{VGG}$}}^{\scalebox{.95}{$\scriptscriptstyle \textrm{DPDD}$}} = \| \phi(\hat{y}_{\scalebox{.95}{$\scriptscriptstyle \textrm{DPDD}$}}) - \phi(y_{\scalebox{.95}{$\scriptscriptstyle \textrm{DPDD}$}})\|_1\\
\label{eq:loss:dp}
\end{equation}
Please note that the losses are applied to all scales. We will present in Sec. \ref{subsec:Analysis_and_Discussion} that our training strategy contributes significantly to the final restoration performance.
\section{Experiments}
\label{sec:experiment}
\subsection{Datasets \& Implementation}
\paragraph{Datasets} We perform experiments on five publicly available datasets for defocus deblurring evaluation, including CUHK \cite{shi2014discriminative}, DPDD \cite{abuolaim2020defocus}, LFDOF \cite{ruan2021aifnet}, PixelDP\cite{abuolaim2020defocus} and RealDOF \cite{lee2021iterative}, as shown in Tab. \ref{tab:dataset_summary}. Specifically, being different from DPDD which is collected with different apertures in consecutive shots, RealDOF is captured by a customized dual-camera system with two Sony $\alpha$7R IV cameras, which are attached to a vertical rig with a beam splitter and equipped with a multi-camera trigger for simultaneous capturing. The images are then post-processed for geometric and photometric alignments. 
\begin{table}[htb] \scriptsize
    \setlength{\abovecaptionskip}{0.05cm}
    \setlength{\belowcaptionskip}{0.05cm}
	\begin{center}
		\begin{tabular}{@{}p{1.4cm}P{0.8cm}P{2cm}p{2.8cm}@{}}
			\toprule
			Dataset   & \# Image & Resolution & Collect Method  \\ 
			\midrule
			CUHK \cite{shi2014discriminative}  & 704 &  $\sim{470}\times610$ & Internet \\
			DPDD \cite{abuolaim2020defocus}  & 500 & $1120\times1680$ &  Canon EOS 5D Mark IV  \\ 
			LFDOF \cite{ruan2021aifnet} & 12k & $688\times1008$ & Lytro Illum   \\       
			PixelDP \cite{abuolaim2020defocus}  & 13 & $\sim{1680}\times1120$ & Google Pixel 4 Smartphone  \\  
			RealDOF \cite{lee2021iterative} & 50 & $\sim{1536}\times2320$ & Sony $\alpha$7R IV \\ 
			\bottomrule
		\end{tabular}
	\end{center}
	\caption{Datasets adopted for training and testing.}
	\label{tab:dataset_summary}
\end{table}
Note that CUHK and PixelDP have no all-in-focus ground truth, as the former one is collected from the Internet while the latter is due to the fixed aperture of smartphone. We utilize LFDOF and DPDD for training and the remaining datasets for evaluation. 

\vspace{0.08cm}
\noindent \textbf{Implementation} { } We implement and evaluate our models using PyTorch with Tesla V100-32GB. We use the Rectified-Adam optimizer \cite{liu2019variance} with $\beta_{1} =0.9$ and $\beta_{2} =0.99$. The initial learning rate is set to $10^{-4}$ when trained on LFDOF for 200 epochs and $10^{-5}$ trained on DPDD for 100 epochs. The updating strategy of learning rate is similar to that of Zhu \etal \cite{zhu2017unpaired}, where the same learning rate is used for the first 100 epochs and then decayed linearly to zero over the rest 100. We set the batch size to 8 and patch size to $320 \times 320$ augmented with Gaussian noise, gray-scale image conversion and scaling. 

\subsection{Comparison to the State-of-the-Art Methods}
\label{subsec:Comparison_to_the_State-of-the-art}
\begin{table*}[htb] \scriptsize
    \setlength{\abovecaptionskip}{0.05cm}
    \setlength{\belowcaptionskip}{0.05cm}
	\begin{center}
		\begin{tabular}{p{1.6cm}p{1.4cm}p{1.4cm}p{1.4cm}p{1.4cm}p{1.4cm}p{1.4cm}p{1.4cm}p{1.4cm}p{1.4cm}}
			\toprule
			\multirow{2}{*}[-0.45em]{Method} & \multicolumn{3}{c}{DPDD Dataset} & \multicolumn{3}{c}{RealDOF Dataset} & \multirow{2}{*}[-0.45em]{Params (M)} \\
			\cmidrule(rl){2-4}                      \cmidrule(l){5-7}  
			&\hfil PSNR$\uparrow$ &\hfil SSIM$\uparrow$ &\hfil LPIPS$\downarrow$   &\hfil PSNR$\uparrow$  &\hfil SSIM$\uparrow$  &\hfil LPIPS$\downarrow$  &\hfil    \\ 
			\midrule
			
			Input        & \hfil 23.890 & \hfil 0.725 &\hfil  0.349  &\hfil  22.333 &\hfil  0.633    &\hfil  0.524 &\hfil - \\ 
			\midrule
		        
			 DPDNet$_S$ \cite{abuolaim2020defocus} & \hfil 24.388 & \hfil 0.749 &\hfil 0.277   &\hfil  22.870 &\hfil  0.670    &\hfil  0.425 &\hfil 31.03\\ 
			 AIFNet \cite{ruan2021aifnet}        & \hfil 24.213 & \hfil 0.742 &\hfil  0.309  &\hfil  23.093 &\hfil  0.680    &\hfil  0.413 &\hfil 41.55\\   
			 IFANet\cite{lee2021iterative}     & \hfil 25.366 & \hfil 0.789 &\hfil 0.217   &\hfil  24.709 &\hfil  0.749   &\hfil  0.306  &\hfil 10.48\\ 
			 KPAC \cite{son2021single}     & \hfil 25.221 & \hfil 0.774 &\hfil  0.226   &\hfil  23.984 &\hfil  0.716    &\hfil  0.336  &\hfil 2.06\\ 	
			\midrule
			Ours          &\hfil \textbf{25.725} &\hfil \textbf{0.791} &\hfil \textbf{0.183}  &\hfil \textbf{25.745} &\hfil \textbf{0.771}    &\hfil \textbf{0.257}  &\hfil 11.69\\
			\bottomrule
		\end{tabular}
	\end{center}
	\caption{Quantitative comparison between our network against existing learning-based methods on single image defocus deblurring. Both datasets are evaluated using the codes and trained weights provided by the authors.}
	\label{tab:comparison_with_others}
    \vspace{-0.2cm}
\end{table*}
\begin{figure*}[htb] 
    \setlength{\abovecaptionskip}{0.1cm}
    \setlength{\belowcaptionskip}{0.1cm}
	\centering
	\includegraphics[width=0.95\linewidth]{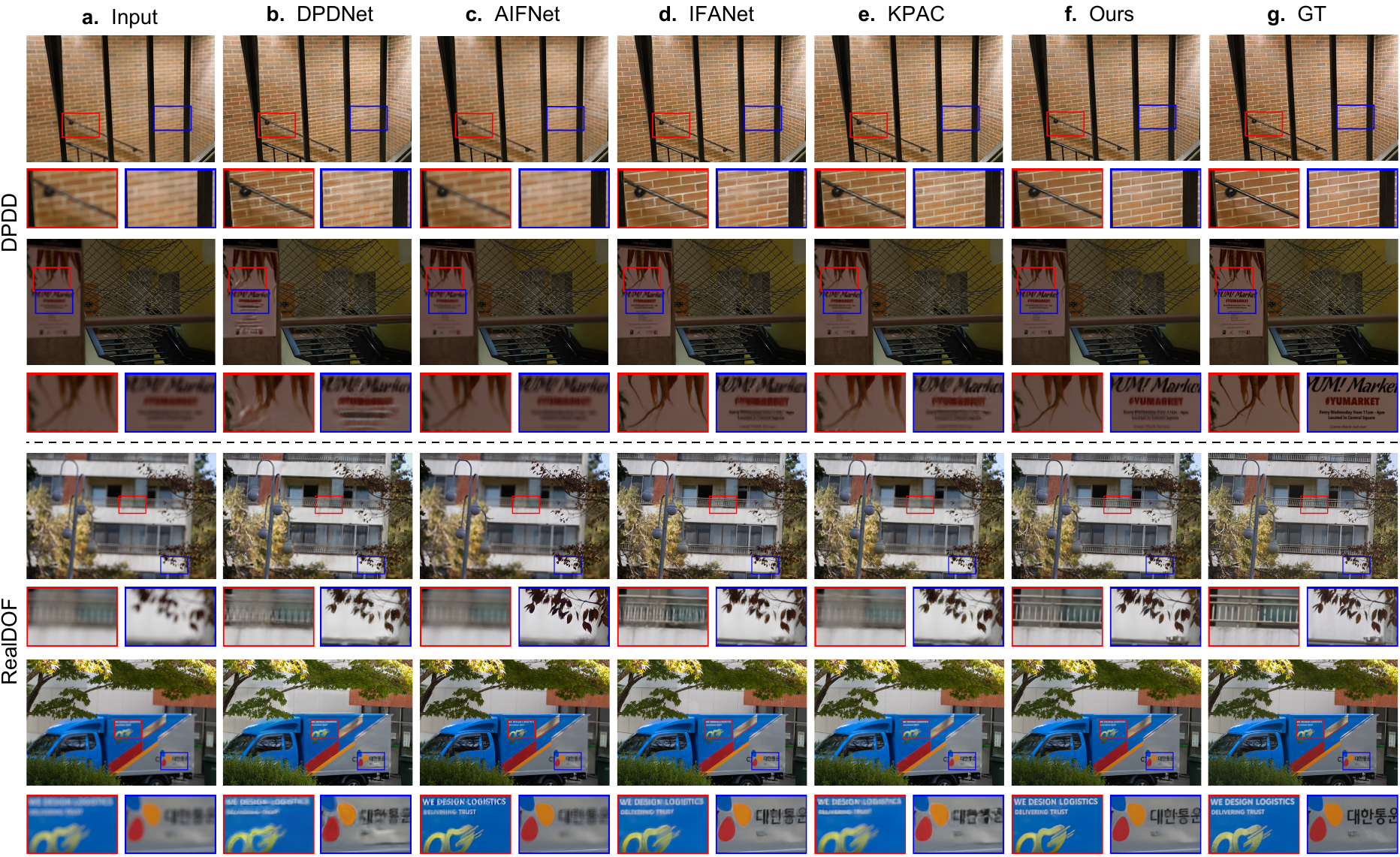}
	\caption{Qualitative evaluation on DPDD and RealDOF datasets among DPDNet$_S$ \cite{abuolaim2020defocus}, AIFNet \cite{ruan2021aifnet}, IFANet\cite{lee2021iterative}, KPAC \cite{son2021single} and ours.  }
	\label{fig:comparison_with_others}
	\vspace{-0.5cm}
\end{figure*}
\paragraph{Evaluation} We compare our proposed method with the four latest learning-based single image defocus deblurring works: DPDNet$_S$ \cite{abuolaim2020defocus}, AIFNet \cite{ruan2021aifnet}, IFANet \cite{lee2021iterative}, and KPAC \cite{son2021single}. All these networks are trained on the DPDD dataset\cite{abuolaim2020defocus} except AIFNet which is trained on LFDOF. Specifically, IFANet needs to incorporate dual views in their network training to estimate the disparity map, while others do not. We use the codes and pre-trained weights released by the authors for comparison, then further evaluate their performance using RealDOF dataset \cite{lee2021iterative}.

In Tab. \ref{tab:comparison_with_others}, we report the quantitative result using three standard evaluation metrics, including  Peak Signal-to-Noise Ratio (PSNR), Structural Similarity (SSIM) \cite{wang2004image} and Learned Perceptual Image Patch Similarity (LPIPS) \cite{zhang2018unreasonable}. We also list the parameter numbers for readers' reference. Our proposed method performs the best among all by a significant margin especially on the RealDOF test set. For instance, our network outperforms DPDNet$_S$, AIFNet, IFANet and KPAC by 12.6\% (2.88dB), 11.5\% (2.65dB), 4.2\% (1.04dB) and 7.3\% (1.76dB) in terms of PSNR respectively. Please note that RealDOF is not used for training, thus it is a fairer benchmark test set for comparing network performance.

Figure \ref{fig:comparison_with_others} shows the corresponding qualitative comparison. Although DPDNet$_S$ can reduce the defocus blur to some extent, it produces artifacts as shown in the second row in Fig. \ref{fig:comparison_with_others}b ). AIFNet gives sharp details in some cases, for instance, the leaves in third row in Fig. \ref{fig:comparison_with_others}c and the characters in fourth row, but it fails to remove defocus blur in other cases. This can be explained by their two-step network architecture, in which the final performance is partially determined by the intermediate defocus map. Thus, AIFNet fails to restore high quality details when the estimated defocus map is inaccurate. KPAC performs slightly better than DPDNet$_S$, while in general it gives limited performance due to its small model capacity. Despite that IFAN shows competitive restoration performance, our proposed method performs better in restoring text (second and fourth row in Fig. \ref{fig:comparison_with_others}f), texture (first row in Fig. \ref{fig:comparison_with_others}f) and object boundary (third row in Fig.  \ref{fig:comparison_with_others}f). It is worth noting that IFAN needs dual views for network training while ours only needs single view.
More results regarding the visual comparison, model complexity, computational costs, \etc are presented in the supplementary material.

\begin{figure*}[htb] 
    \setlength{\abovecaptionskip}{0.1cm}
    \setlength{\belowcaptionskip}{0.1cm}
	\centering
	\includegraphics[width=0.95\linewidth]{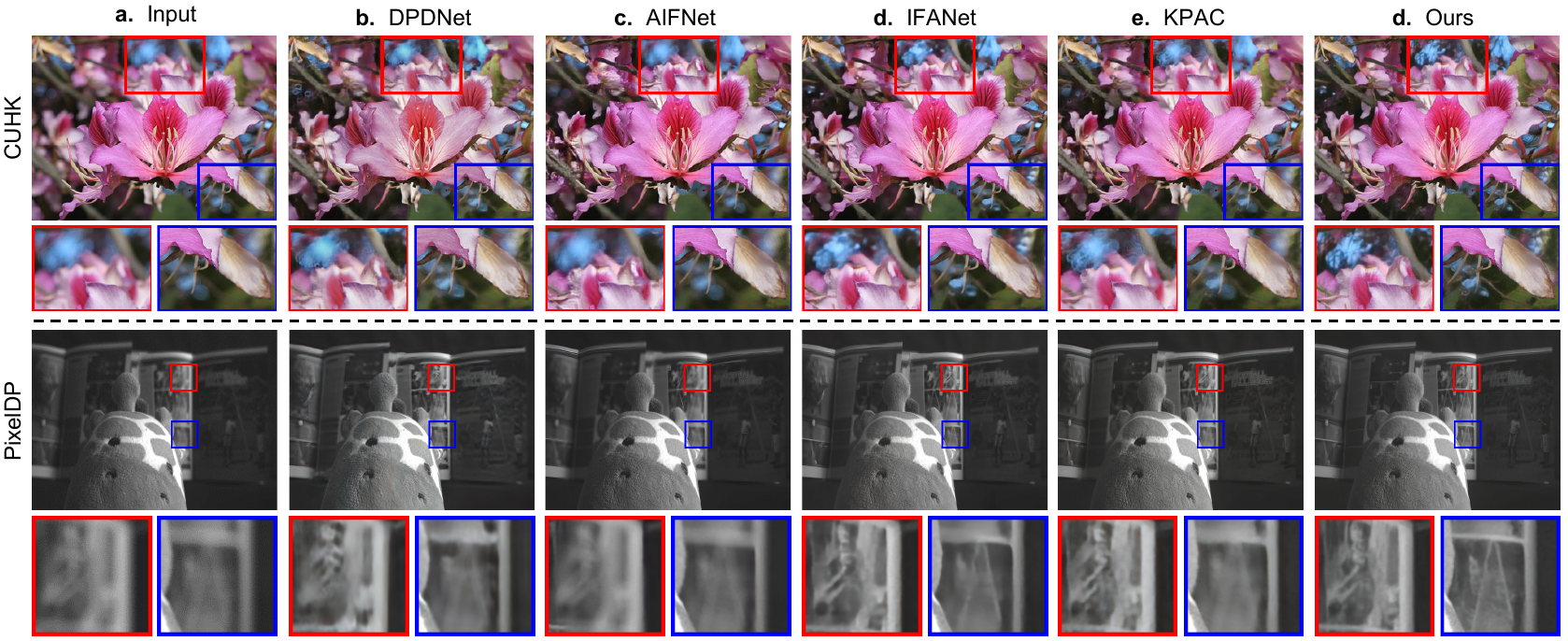}
	\caption{Qualitative comparison among DPDNet$_S$\cite{abuolaim2020defocus}, AIFNet \cite{ruan2021aifnet}, IFAN \cite{lee2021iterative}, KPAC \cite{son2021single} and ours. Image in the first row is from CUHK \cite{shi2014discriminative} and that in the second row is from PixelDP \cite{abuolaim2020defocus}. No all-in-focus ground truth is provided in these two datasets.}
	\label{fig:comparison_no_ref}
	\vspace{-0.4cm}
\end{figure*}

\vspace{0.08cm}
\noindent \textbf{Generalization Ability} { } To inspect the generalization ability of our network, we further compare the visual performance of the networks using CUHK \cite{shi2014discriminative} and PixelDP \cite{abuolaim2020defocus} datasets. CUHK targets for blur detection with relatively small spatial resolution, and all images are collected from the Internet thus no all-in-focus ground truth are provided. PixelDP is collected with Google Pixel 4 smartphone which has a fixed aperture size and the image data is limited to one of the Green channels in the ray-Bayer frame. Figure \ref{fig:comparison_no_ref} reports the visual comparison among the five networks and a similar conclusion can be drawn: our method gives the best generalization ability and can successfully restore the fine details regardless of the camera type. More results are provided in the supplementary material.
\subsection{Analysis and Discussion}
\label{subsec:Analysis_and_Discussion}
\noindent{In} this section, we conduct comprehensive ablation studies and analysis.

\vspace{0.08cm}
\noindent \textbf{Why LFDOF?}{ } To understand the necessity of LFDOF for defocus deblurring, we train our network on LFDOF only, DPDD only and both. Here, we use `LFDOF \& DPDD' to represent the network trained on LFDOF then fine-tuned on DPDD for convenience. Table \ref{tab:trained_on_dpd_lfdof} and Fig. \ref{fig:train_on_lf_dpd} show the quantitative and qualitative results evaluated using RealDOF test set. It is observed that our network trained on DPDD gives better scores in terms of all metrics than LFDOF. This can be explained by the domain difference between the light field-generated and real defocused images. However, the visual performance is not in line with the quantitative performance. The network trained on LFDOF produces sharper content and details than the one trained on DPDD does owing to the accurate pixel correspondence of LFDOF. However, at the same time, it also introduces artifacts (see wall in Fig. \ref{fig:train_on_lf_dpd}) due to the defocus blur discrepancy between light field generated and real data. Our strategy to train the network on LFDOF then fine-tune on DPDD largely outperforms the networks trained on either dataset alone and generates the best quantitative (increase up to 11.57\% and 4.23\% in terms of PSNR) and qualitative results.
\begin{table}[htb] \scriptsize
    \setlength{\abovecaptionskip}{0.05cm}
    \setlength{\belowcaptionskip}{0.1cm}
	\begin{center}
		\begin{tabular}{@{}p{2.6cm}p{1.0cm}p{1.0cm}p{1.0cm}p{1.0cm}}
		\toprule
			 \hfil Training Dataset &\hfil PSNR$\uparrow$  &\hfil SSIM$\uparrow$   &\hfil LPIPS$\downarrow$  \\ 
			\midrule
			
			\hfil LFDOF \cite{ruan2021aifnet}&\hfil  23.076  &\hfil 0.698  &\hfil 0.378    \\          
			
			\hfil DPDD \cite{abuolaim2020defocus} &\hfil  24.700  &\hfil 0.744 &\hfil 0.337    \\
			\midrule
			\hfil LFDOF \& DPDD (Ours) &\hfil  \textbf{25.745}  &\hfil \textbf{0.771}  &\hfil \textbf{0.257}   \\   
			\bottomrule
		\end{tabular}
	\end{center}
	\caption{Quantitative comparison of the proposed network trained on LFDOF only, DPDD only and both datasets. The results are tested on RealDOF test set. }
	\label{tab:trained_on_dpd_lfdof}
	\vspace{-0.1cm}
\end{table}
\begin{figure}[htb]
    \setlength{\abovecaptionskip}{0.05cm}
    \setlength{\belowcaptionskip}{0.1cm}
	\centering
	\includegraphics[width=0.95\linewidth]{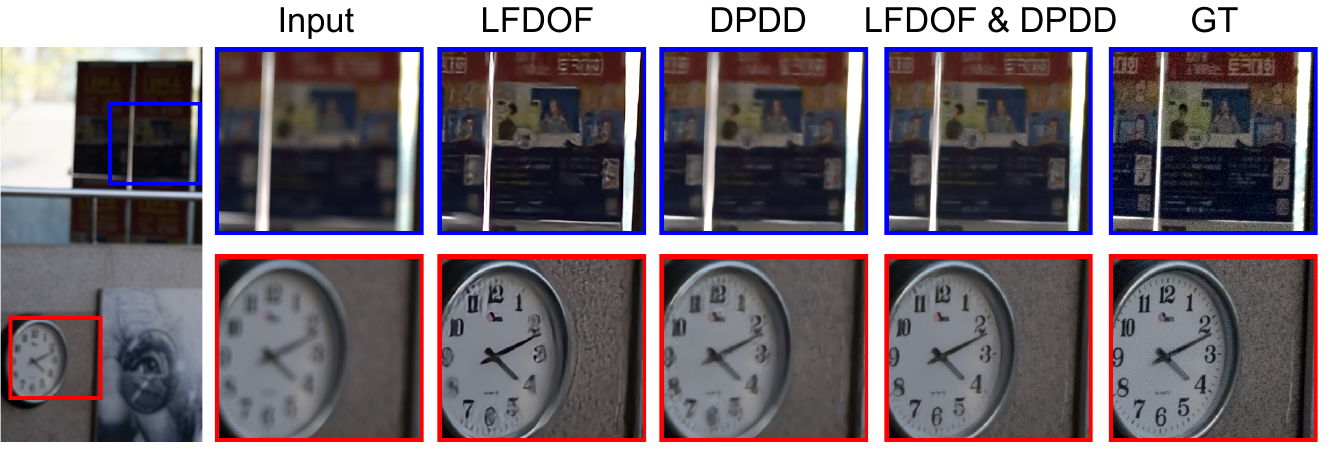}
	\caption{Visual comparison of our network trained on LFDOF only, DPDD only and both.}
	\label{fig:train_on_lf_dpd}
	\vspace{-0.1cm}
\end{figure}

\vspace{0.08cm}
\noindent \textbf{Performance Gain}{ } To further validate the proposed training strategy, we have retrained two state-of-the-art networks, DPDNet$_S$ \cite{abuolaim2020defocus} and KPAC \cite{son2021single}, to see whether our training scheme could improve the performance of these two networks. AIFNet and IFANet are not listed here because the former requires defocus map as the ground truth and the latter requires dual views for training. For KPAC, we choose the 3-level model with 2 KPAC blocks for comparison. Except the loss and learning rate which are set to be the same as ours in the two training stages, other parameter settings for DPDNet$_S$ and KPAC are the same as their original configurations. The network performance are evaluated on 76 test samples in DPDD dataset. Table \ref{tab:performance_gain} shows that DPDNet$_S$, KPAC and our proposed architecture have gained 0.511dB, 0.249dB, and 0.253dB in terms of PSNR when networks are trained on LFDOF \& DPDD. The visual quality has also been largely improved as shown in Fig. \ref{fig:performance_gain}.
\begin{table}[htb] \scriptsize
	 \setlength{\abovecaptionskip}{0.05cm}
    \setlength{\belowcaptionskip}{0.05cm}
	\begin{center}
		\begin{tabular}{@{}p{1.5cm}p{0.6cm}p{0.6cm}p{0.6cm}p{0.6cm}p{0.6cm}p{0.6cm}}
		\toprule
		\multirow{2}{*}[-0.45em]{ Method} & \multicolumn{3}{c}{DPDD} & \multicolumn{3}{c}{LFDOF \& DPDD} \\ 
			\cmidrule(rl){2-4}                      \cmidrule(l){5-7}
			 & PSNR$\uparrow$  & SSIM$\uparrow$ & LPIPS$\downarrow$  & PSNR$\uparrow$  & SSIM$\uparrow$ & LPIPS$\downarrow$  \\ 
			\midrule
			
			 DPDNet$_S$ \cite{abuolaim2020defocus} &  24.388  & 0.749 & \textbf{0.277} & \textbf{24.899}  & \textbf{0.761} & 0.278   \\          
			 KPAC \cite{son2021single} &  25.221  & 0.774 &0.226 & \textbf{25.470}  & \textbf{0.780} & \textbf{0.220}   \\ 
			 Ours &  25.472  & 0.787 & 0.246  & \textbf{25.725} & \textbf{0.791}& \textbf{0.183}   \\     
			\bottomrule
		\end{tabular}
	\end{center}
	\caption{Performance gain of DPDNet$_S$ \cite{abuolaim2020defocus}, KPAC \cite{son2021single} and ours when trained on LFDOF \& DPDD. The networks are tested on 76 test samples in DPDD dataset.}
	\label{tab:performance_gain}
	\vspace{-0.1cm}
\end{table}
\begin{figure}[htb]
    \setlength{\abovecaptionskip}{0.1cm}
    \setlength{\belowcaptionskip}{0.1cm}
	\centering
	\includegraphics[width=0.95\linewidth]{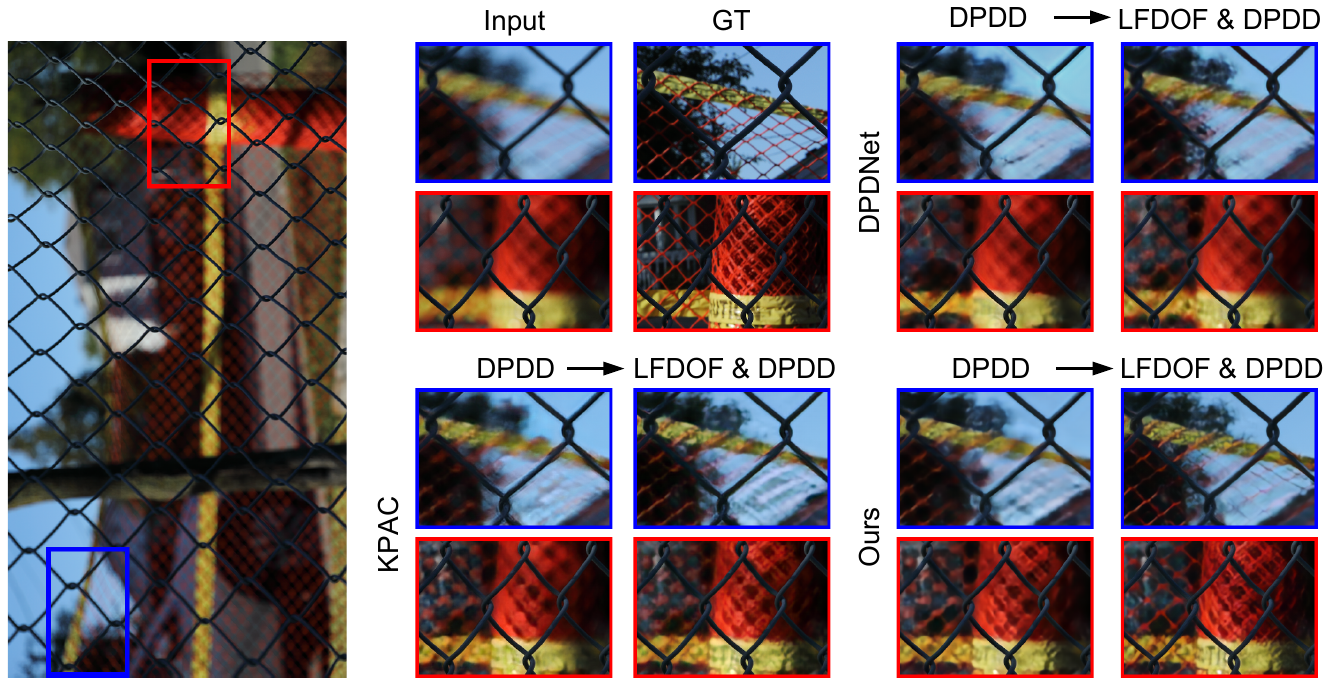}
	\caption{Qualitative comparison of network performance using DPDD only (left column) and LFDOF and DPDD datasets (right column) for training. All networks get improved performance. }
	\label{fig:performance_gain}
\end{figure}

\noindent\textbf{Loss} { }Our proposed method trains the network on LFDOF with $\mathcal{L}_1$ loss, then on DPDD with feature loss. One may wonder if it is possible to (1) directly mix the datasets for training instead of using a two-stage training strategy, and (2) adopt the same $\mathcal{L}_1$ loss in the fine-tuning stage. To answer these questions, we ablate the use of loss, then evaluate their performance on DPDD and RealDOF datasets as shown in Tab. \ref{tab:loss_ablate} and Fig. \ref{fig:loss_ablate}. To balance the ratio of the two datasets, we augment DPDD dataset (350 images) for 32 times, then mix with LFDOF (11261 images). The network performs similarly when using $\mathcal{L}_1$ loss alone and $\mathcal{L}_1$ combined with $\mathcal{L}_{\scalebox{.9}{$\scriptscriptstyle \textrm{VGG}$}}$ loss on the mixed dataset, while performing slightly worse than the version trained on DPDD only (see Tab. \ref{tab:trained_on_dpd_lfdof} and Tab. \ref{tab:performance_gain}). When adopting the same $\mathcal{L}_1$ loss in two stages, the quantitative result in terms of PSNR and SSIM is comparable with our final one, while the perceptual score LPIPS is relatively worse on DPDD dataset. It is because the per-pixel loss on DPDD may not lead to an optimal performance, even the misalignment is not obvious for human perception (\eg, Fig. \ref{fig:cross_correlation}) but sensitive to the network. In addition, its performance on RealDOF is less favorable with 0.62dB lower than our final one in terms of PSNR. In Fig. \ref{fig:loss_ablate}, the results in red frames are produced by our final model, which yield the most realistic and fine details. Both quantitative and qualitative performance further validate and support the proposed training strategy.
\begin{table}[htb] \scriptsize

    \setlength{\abovecaptionskip}{0.1cm}
    \setlength{\belowcaptionskip}{0.1cm}
    \centering
	\resizebox{\linewidth}{!}{%
		\begin{tabular}{@{}p{0.6cm}p{0.2cm}p{0.55cm}p{0.35cm}p{0.45cm}p{0.45cm}p{0.45cm}p{0.45cm}p{0.45cm}p{0.45cm}}
		\toprule
		\multirow{2}{*}[-0.45em]{Strategy}&\multirow{2}{*}[-0.45em]{ No.}& \multicolumn{2}{c}{Loss}& \multicolumn{3}{c}{DPDD} & \multicolumn{3}{c}{RealDOF} \\ 
		\cmidrule(rl){3-4}  	\cmidrule(rl){5-7}                      \cmidrule(l){8-10}
			
			 & &S$_1$ &S$_2$& PSNR$\uparrow$  & SSIM$\uparrow$  & LPIPS$\downarrow$  & PSNR$\uparrow$ & SSIM$\uparrow$ & LPIPS$\downarrow$ \\ 
			\midrule
		
			 \multirow{2}{*}[0em]{Mix} & \hfil \textbf{a} &
			 $\mathcal{L}_1$ & - & 25.439  & 0.793 & 0.237  & 24.634 & 0.751 & 0.330    \\   
			  &\hfil \textbf{b}&  \resizebox{0.9cm}{!}{$\mathcal{L}_1\!+\! \lambda\mathcal{L}_{\scalebox{.9}{$\scriptscriptstyle \textrm{VGG}$}}$} & - & 25.469  & 0.793 & 0.236  & 24.684 & 0.751 &0.329    \\ 
			\midrule  
			 \multirow{2}{*}[0em]{Fine-tune} &\hfil \textbf{c}& $\mathcal{L}_1$ &$\mathcal{L}_1$ &  \textbf{25.755}  & \textbf{0.797} & 0.232  & 25.130 & 0.768 & 0.310  \\ 
			 &\hfil \textbf{d}& $\mathcal{L}_1$ &{$\mathcal{L}_{\scalebox{.9}{$\scriptscriptstyle \textrm{VGG}$}}$} & 25.725  & 0.791 & \textbf{0.183}  & \textbf{25.745}  & \textbf{0.771} & \textbf{0.257}  \\

			\bottomrule
		\end{tabular}}

	\caption{Quantitative comparison of the training strategy with respect to dataset and loss. S$_1$ and S$_2$ represent the main training on LFDOF and fine-tune on DPDD respectively. There is only one stage when training on the mixed dataset and $\lambda$ is set to $10^{-5}$.}
	\label{tab:loss_ablate}
	\vspace{-0.1cm}
\end{table}

\begin{figure}[htb]
    \setlength{\abovecaptionskip}{0.1cm}
    \setlength{\belowcaptionskip}{0.1cm}
	\centering
	\includegraphics[width=0.9\linewidth]{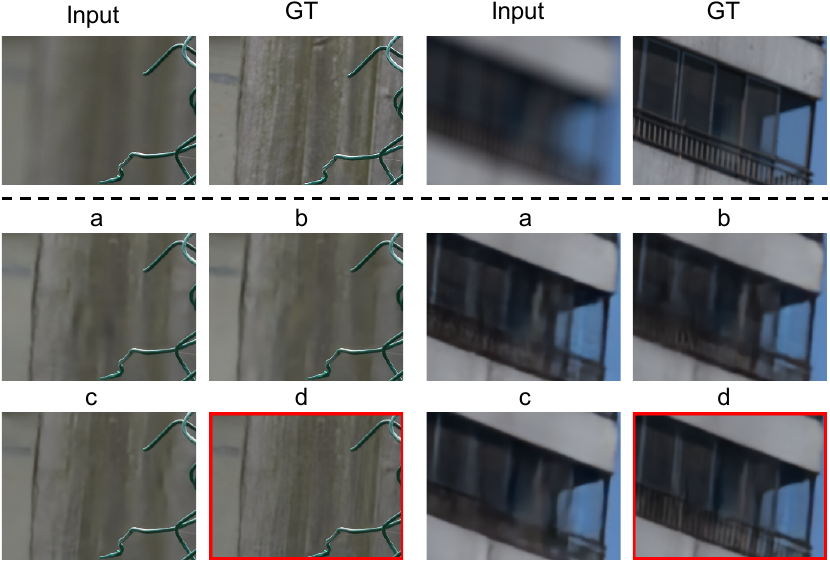}
	\caption{Visual comparison of the training strategy with respect to dataset and loss. Here a, b, c, and d indicate the corresponding training strategies in Tab. \ref{tab:loss_ablate}.}
	\label{fig:loss_ablate}
\end{figure}

\vspace{0.08cm}
\noindent \textbf{DRB Configurations}{ } To validate the effect of each component in DRB, we conduct an ablation study and report the results in Tab. \ref{tab:block_design} and Fig. \ref{fig:block_design}. We compare our final model with four variant networks: removing one component each time and resulting in the block with ($\hat{x}_{\downarrow_n}$ , $\Delta \hat{x}_{\downarrow_n}$, - ), ( - , $\Delta \hat{x}_{\downarrow_n}$, $\Delta r_{\downarrow_n}$), ($\hat{x}_{\downarrow_n}$, - , $\Delta r_{\downarrow_n}$), and baseline that direct outputs the restored result without DRB. Those variants are all with multi-scale architecture. Another variation with one full DRB (remove $\mathcal{R}_{\downarrow_{\{8, 4, 2\}}}$ thus leaving $\mathcal{R}$ only) is also added for comparison. Both quantitative and qualitative result show the final model is able to restore the finest details as shown in Tab. \ref{tab:block_design} and Fig. \ref{fig:block_design}. For visual quality, Figure \ref{fig:block_design} demonstrates that only the DRB with all the components is capable of restoring realistic details (see last two in the second row).

\begin{table}[htb] \scriptsize
    \setlength{\abovecaptionskip}{0.1cm}
    \setlength{\belowcaptionskip}{0.1cm}
	\begin{center}
	\resizebox{\linewidth}{!}{%
		\begin{tabular}{@{}p{0.5cm}p{1.0cm}p{0.5cm}p{1.2cm}p{0.7cm}p{0.7cm}p{0.7cm}}
		\toprule
			$\hat{x}_{\downarrow_n}$ & $\Delta \hat{x}_{\downarrow_n}$  & $\Delta r_{\downarrow_n}$  & multi-scale & PSNR$\uparrow$ & SSIM$\uparrow$ & LPIPS$\downarrow$\\
			\midrule
			\hfil  &\hfil baseline    &\hfil  &  &\hfil 25.327 &\hfil 0.749  &\hfil 0.285  \\ 			
			\hfil \checkmark &\hfil  \checkmark  &\hfil  & \hfil \checkmark &\hfil 25.539 &\hfil 0.763  &\hfil 0.271  \\          
			\hfil  &\hfil  \checkmark  &\hfil \checkmark  &\hfil \checkmark &\hfil 25.576 &\hfil 0.763  &\hfil 0.267 \\
			\hfil \checkmark &\hfil    &\hfil \checkmark&\hfil \checkmark&\hfil 25.515  &\hfil 0.765 &\hfil 0.274 \\ 
			
				\hfil \checkmark &\hfil  \checkmark  &\hfil \checkmark &\hfil  &\hfil 25.532  &\hfil 0.757 &\hfil 0.272 \\

			\hfil \checkmark &\hfil  \checkmark  &\hfil \checkmark &\hfil \checkmark  &\hfil \textbf{25.745}  &\hfil \textbf{0.771} &\hfil \textbf{0.257} \\   			
			\bottomrule
		\end{tabular}}
	\end{center}
	\caption{Ablation study on each component in DRB and the multi-scale restoration strategy. Performance is evaluated on RealDOF.}
	\label{tab:block_design}
	\vspace{-0.3cm}
\end{table}
\begin{figure}[htb]
    \setlength{\abovecaptionskip}{0.1cm}
    \setlength{\belowcaptionskip}{0.1cm}
	\centering
	\includegraphics[width=.9\linewidth]{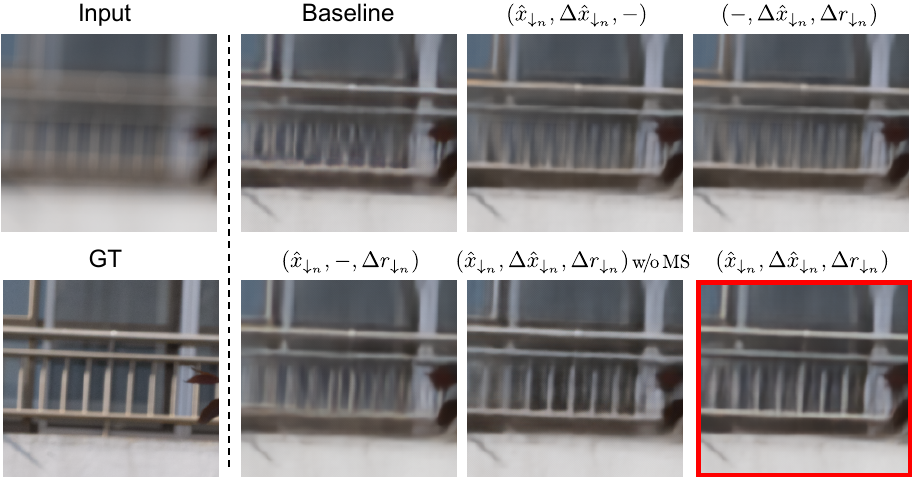}
	\caption{Qualitative results of the ablation study of DRB configurations evaluated on RealDOF. }
	\label{fig:block_design}
	\vspace{-0.3cm}
\end{figure}

\vspace{0.08cm}
\noindent\textbf{Dynamic Kernel Size} { }We analyze the effect of dynamic kernel size in our restoration performance. As illustrated in Tab. \ref{tab:kernel_size_ablate}, we experimentally identify the best performing kernel size for DPDD dataset to be 9 while that for RealDOF to be 7. Considering the overall performance and parameter number, kernel size 7 is chosen. 
\begin{table}[htb] \scriptsize
    \setlength{\abovecaptionskip}{0.1cm}
    \setlength{\belowcaptionskip}{0.1cm}
	\begin{center}
		\begin{tabular}{@{}p{1.0cm}p{0.8cm}p{0.8cm}p{0.8cm}p{0.8cm}p{0.8cm}p{0.8cm}@{}}
		\toprule
		\multirow{2}{*}[-0.45em]{\hfil Kernel Size} & \multicolumn{3}{c}{DPDD} & \multicolumn{3}{c}{RealDOF} \\ 
			\cmidrule(rl){2-4}                      \cmidrule(l){5-7}
			
			 &\hfil PSNR$\uparrow$  &\hfil SSIM$\uparrow$  &\hfil LPIPS$\downarrow$ &\hfil PSNR$\uparrow$  &\hfil SSIM$\uparrow$ &\hfil LPIPS$\downarrow$ \\ 
			\midrule
			
			\hfil 5  &\hfil  25.660  &\hfil 0.789 &\hfil 0.185 &\hfil 25.564  &\hfil 0.765 &\hfil 0.271     \\          
			\hfil 7 &\hfil  25.725  &\hfil 0.791&\hfil 0.183 &\hfil \textbf{25.745}  &\hfil \textbf{0.771} &\hfil \textbf{0.257}   \\ 
			\hfil 9 &\hfil  \textbf{25.752}  &\hfil 0.790 &\hfil \textbf{0.182} &\hfil 25.552  &\hfil 0.764 &\hfil 0.265  \\
			\hfil 11 &\hfil  25.716  &\hfil \textbf{0.793}  &\hfil 0.183 &\hfil 25.625  &\hfil 0.768 &\hfil 0.265  \\ 
			\bottomrule
		\end{tabular}
	\end{center}
	\caption{Quantitative evaluation on DPDD and RealDOF datasets with respect to the kernel size in DRB block.}
	\label{tab:kernel_size_ablate}
\end{table}

\vspace{0.08cm}
\noindent\textbf{AIFNet \vs Ours} { } Both AIFNet and ours are trained on LFDOF dataset. However, AIFNet employs defocus map estimation network followed by deblurring network while our network is an end-to-end architecture without explicit defocus map estimation. In order to compare these two networks, we train and test their performance only on LFDOF. Table \ref{tab:aifnet_vs_ours} shows that our network architecture outperforms AIFNet by 0.726dB in terms of PSNR without the help of defocus map. Accurate defocus map may not be able to boost the restoration performance, whereas inaccurate ones will limit or impede the performance. This further validates the effectiveness of our proposed network architecture.

\begin{table}[htb] \scriptsize
    \setlength{\abovecaptionskip}{0.05cm}
    \setlength{\belowcaptionskip}{0.05cm}
	\begin{center}
		\begin{tabular}{@{}p{1.4cm}p{1.0cm}p{1.0cm}p{1.0cm}}
		
		\toprule
		\multirow{2}{*}[-0.45em]{ Method}& \multicolumn{3}{c}{Evaluation on LFDOF}  \\ 
			\cmidrule(rl){2-4}                      
			
			   & PSNR$\uparrow$  & SSIM$\uparrow$  & LPIPS$\downarrow$  \\ 
			\midrule
			
			AIFNet  &  29.677  & 0.884 & 0.202      \\   
			Ours & \textbf{30.403}  & \textbf{0.891} & \textbf{0.145}      \\ 

			\bottomrule
		\end{tabular}
	\end{center}
	
	\caption{Quantitative comparison between AIFNet and our network evaluated on 725 images from LFDOF test set. Both networks are trained on LFDOF training set.}
	\label{tab:aifnet_vs_ours}
    \vspace{-0.4cm}
\end{table}


\section{Conclusion}
\label{sec:conclusion}
We have proposed a novel method drawing on the synthetic aperture and refocusing features of light fields along with real captured defocus blur dataset to address the single image defocus deblurring problem. Our end-to-end neural network equipped with dynamic residual block is proven to be effective for removing spatially varying defocus blur. We train our network on light field generated dataset with MAE loss for the superior pixel-wise correspondence, then on real defocus dataset with feature loss to fully utilize the advantages of two types of data. We have proved this training strategy can be applied to improve the performance of several existing learning methods. Extensive comparison and ablation studies have demonstrated the effectiveness of our method, which outperforms others by a significant margin on multiple test sets. 

\noindent \textbf{Limitations}{ } Despite showing competitive performance, our proposed method shares some similar limitations with Lee \etal \cite{lee2021iterative} and Son \etal \cite{son2021single} in handling blur with irregular shapes and defocus blur mixed with object motion. We include these failure cases in the supplementary material. Our future work will take these challenging cases into consideration.

{\small

\bibliographystyle{ieee_fullname}
}

\end{document}